# A Key Distribution Scheme for Sensor Networks Using Structured Graphs

*Invited Paper*

Abhishek Parakh
Computer Science Department
Oklahoma State University
Stillwater OK, USA

Subhash Kak
Computer Science Department
Oklahoma State University
Stillwater OK, USA

*Abstract* – **This paper presents a new key predistribution scheme for sensor networks based on structured graphs. Structured graphs are advantageous in that they can be optimized to minimize the parameter of interest. The proposed approach achieves a balance between the number of keys per node, path lengths, network diameter and the complexity of routing algorithm.**

## I. INTRODUCTION

Data in sensor networks, which are used for data collection, intruder detection, traffic and wildlife monitoring and medical procedure, needs to be cryptographically protected in sensitive applications such as battlefield and commerce. Limited computational power and memory in sensor networks precludes the use of asymmetric key cryptography and restricts the number of encryption/decryption keys that can be stored. As a result, the use of symmetric keys and distribution of keys before deployment becomes an obvious choice.

Key predistribution in sensor networks has been modeled based on the theory of random graphs, as proposed by Eschenauer and Gligor (EG) [1]. The EG model is based on the assumptions that no *a priori* deployment knowledge is available, and a sensor can communicate only with a very small fraction of other sensors. However, these assumptions may not apply to every sensor network. For example, in a sensor network in which all or majority of sensors are within communication range of each other, a random graph may be an inefficient model for key distribution as it requires higher number of edges to guarantee connectivity. As a comparison, in a structured graph, only *two* keys per node are required to guarantee connectivity, *irrespective* of the network size. However, in a random graph with only 10,000 nodes, 15 keys are required to ensure that the network is connected with a probability of 0.99999 [1].

Structured graphs can be optimized with respect to different parameters such as network diameter, the average shortest path length between any two nodes in the network, the clustering coefficient, the number of keys a node is required to store and routing complexity.

Approaches similar to EG, require the use of a key discovery phase as the sensors do not know with which other sensors they share keys. This entails a broadcast of message to every other node after deployment which is followed by a challenge/response phase [1]. Conversely, in our model the keys are distributed in a predetermined fashion and a key discovery phase is not required. Also, each key can be chosen to be unique, ensuring that only the intended node can decrypt the message and no other nodes can listen in on the conversation.

After the key discovery phase [1], a node wanting to communicate with a node with which it does not share an encryption/decryption key perform path discovery. There are several problems with path discovery. Determining the shortest path to every other node on the network requires the knowledge and storage of the complete network topology (or the adjacency matrix of size $N^2$) at the node that wants to discover the shortest path. Since in a random graph the adjacency matrix is not predetermined, every node will have to communicate back and forth to build this matrix. This is an enormous communication burden on the network. In the structured graph approach, we eliminate the path discovery phase and present a predetermined efficient routing algorithm that establishes a worst case path length of $\left\lfloor \frac{\log_2(\frac{N}{8})}{2} \right\rfloor + 2$ and requires the storage of only $O(\log_2 N)$ keys.

Chan, *et. al.*, [2] proposed the use of a $q$-composite scheme in which $q > 1$ keys were to be shared between every pair of nodes in a random graph. Although this provided better resilience than [1], it requires an increase in the number of keys to be stored on the nodes or a decrease in pool size. While the first requirement is a burden on memory, decreasing the pool size increases the probability of the same key being reused between more than two nodes.

Du, *et. al.* [3] and Liu and Ning [4] propose a threshold scheme in which the adversary needs to compromise more than a certain number of nodes to subvert the network. In [6-8] deployment knowledge is used to reduce the number of keys required to share. In the proposed scheme, we only assume that the nodes are within communication range of each other without the need for any specific deployment topology and we see that the capture of a single node only reveals $2\log_2(N)-1$ keys.

## II. THE PROPOSED KEY DISTRIBUTION SCHEME

One may view the elements in fig. 1(a) as nodes in a sensor network arranged in the increasing order of their node IDs (1 to $N$), in a sequence. An arc between node $i$ and node $j$ indicates that the nodes share a common encryption/decryption key and hence can communicate securely. Further, every node is assumed to have similar connections where each connection is bidirectional since the shared key makes it possible to communicate in both directions.

If we say that every node $i$ shares a key with nodes $i+1$, $i+2$, $i+4$, $i+8$, $i+16$, …, $i+\log_2(\frac{N}{2})$ (computed modulo $N$), then the a network with $N$ nodes has $N \times \log_2(N)$ total connections. Further, since every connection is bidirectional, and the network is symmetric, such a key distribution strategy results in $2\log_2(N)-1$ connections for every node (where $N$ is a power of 2). This is illustrated in figures 1(b) for $N=16$. A similar routing table has been used in peer-to-peer data look up protocols [5], but, unlike our case, connections there are unidirectional.

In order to securely communicate with the nodes with which node $i$ does not directly share an encryption key, the message is routed via multiple hops in the following manner (*the routing protocol*):

1. To securely communicate with node $q$, node $i$ first encrypts the message using the encryption key it shares with node $m$ that is closest to the destination and with which it (node $i$) has a direct connection and sends the encrypted message to node $m$.
2. Node $m$ then decrypts the message, and checks if node $q$ is its direct contact. If it is, then node $m$ encrypts the message using the encryption key it shares with node $q$ and sends the message to node $q$ directly. However, if $q$ is not one of $m$'s direct contacts, then node $m$ locates the next node, $l$, that is closest to node $q$, among its direct contacts, and encrypts the message with the encryption key its shares with node $l$ and sends the encrypted message to it.
3. Node $l$ repeats step 2, and so on until the message reaches node $q$.

The *closeness* of node $a$ to node $b$ is determined by computing (Node ID $(a)$ − Node ID $(b)$) mod $N$, where Node ID $(a) \leq$ Node ID $(b)$.

The parameters considered in the proposed model for key distribution are the following:

1. Network diameter.
2. Average shortest path length.
3. Number of keys that a node is required to store. And,
4. Clustering coefficient.

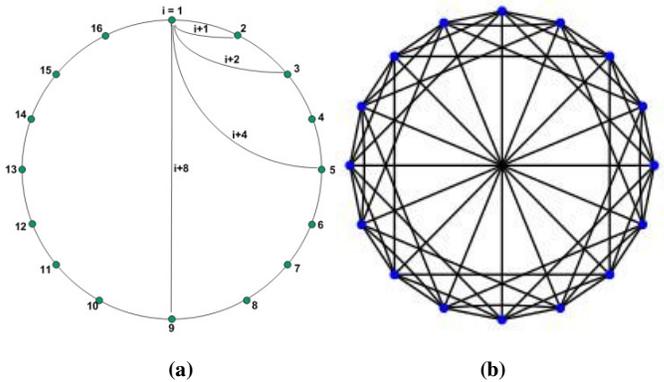

Figure 1. (a) Connections that node 1 establishes. All the nodes establish shared keys in a similar fashion. (b) Connections for all the 16 nodes.

*Theorem 1:* Consider a network with $N$ nodes, where $N = 2^k$, for some integer $k \geq 3$, such that every node shares keys with $2\log_2(N)-1$ other nodes as described above, then any two nodes can securely communicate in at most $\left\lfloor \frac{\log_2(\frac{N}{8})}{2} \right\rfloor + 2$ hops.

*Proof*: In our network, every connection is bidirectional and we wish to determine the upper bound on the distance between any two nodes $i$ and $q$ in the network. Without loss of generality, if node $q$ is not a direct contact of node $i$, then assume that node $i$ contacts one of its direct connections, that is closest to node $q$ and routes the message using the routing algorithm presented above.

Further it is seen that the placement of connections of node $i$ slices the network into pieces such that the largest piece is of size $N/4$. Therefore, node $i$ would, in the worst case, have to contact one of the nodes at the edge of the largest slice. Since, node $i$ sends the message to the connection closest to the destination, the maximum distance after the first hop, between the destination and the node that receives the packet will be $\frac{1}{2} \cdot \frac{N}{4}$. Consequently, the distance after 1-hop is reduced to $M = \frac{N}{8}$ in the worst case.

Routing continues in a similar fashion after the first hop. However, the node connections are such that the largest slice is half the size of the remaining distance, because the remaining part of the network no longer forms a circle. Consequently, the distance between the message and the destination, second hop onwards, only reduces to ¼ the size of the remaining distance. This process is repeated until the destination node is reached by repeatedly dividing the remaining distance by ¼ at each step. Hence, after $m$ steps, the distance from node $q$ is reduced to $\frac{M}{4^m} = \frac{M}{2^{2m}}$.

When $M = 2^x$, where $x =$ even, then repeated division by 4 results in the last step distance being 1; when $x =$ odd, this distance is 2. However, in both cases, the destinatination is one hop away. Consequently, division by 4 stops when either,

$\frac{M}{2^{2m}} = 1 \Rightarrow M = 2^{2m} \Rightarrow m = \frac{\log_2 M}{2}$, when $\log_2 M$ is even, and

$\frac{M}{2^{2m}} = 2 \Rightarrow M = 2^{2m+1} \Rightarrow m = \frac{\log_2 M - 1}{2}$, when $\log_2 M$ is odd.

Adding the first hop and the last hop to the above expression, and noting that when $\log_2 N$ is even $\log_2 M$ is odd and vice versa, we get the following upper bounds for the path lengths,

Path length = $\frac{\log_2 \frac{N}{8}}{2} + 2$, when $\log_2 N$ is odd, and

Path length = $\frac{\log_2 \frac{N}{8} - 1}{2} + 2$, when $\log_2 N$ is even.

Combining the two cases above yields an expression for the upper bound on the path length: $\left\lfloor \frac{\log_2 \frac{N}{8}}{2} \right\rfloor + 2$. ∎

*Example 1.* For a network of size $N = 1024$, the diameter of the network is $\left\lfloor \frac{\log_2 \frac{1024}{8}}{2} \right\rfloor + 2 = \left\lfloor \frac{7}{2} \right\rfloor + 2 = 5$. Whereas, when $N = 2048$, the diameter of the network is $\left\lfloor \frac{\log_2 \frac{2048}{8}}{2} \right\rfloor + 2 = \left\lfloor \frac{8}{2} \right\rfloor + 2 = 6$.

One can argue that for secure communication between sensor nodes, the nodes must share a key, represented by a link between the nodes. Each node has a "*view*" that is limited to these direct links. For the purpose of secure transmission of messages, a node can effectively only "*see*" its direct links and to send a message to a node other than its direct links, the message is routed via intermediate nodes, encrypting and decrypting the data at each hop.

*Theorem 2*: The routing protocol described above, results in the shortest possible path, for the given connections, between the starting node and the destination node.

*Proof*: Since every node's "*view*" is limited to $2\log_2(N) - 1$ nodes, it can securely only contact those nodes that are its direct connections. Hence, the best strategy without a global *view* (knowledge) is to use a greedy approach, i.e. to choose the local best at every node.

To securely route a message to the destination node, every node sends the message to one of its direct contact that is closest to the destination node. As a consequence, the distance between the starting node and the destination node decreases at each step to the smallest possible at that step, making the algorithm optimal, for the given connections, resulting in the shortest possible path. ∎

The *diameter* of a network is defined as the longest shortest path length in the network. From theorem 1 and 2 the diameter of the network is $\left\lfloor \frac{\log_2 \frac{N}{8}}{2} \right\rfloor + 2$, when $N$ is a power of 2. And when $2^k < N < 2^{k+1}$, for some integer $k$, the diameter varies between the diameters for network of sizes $2^k$ and $2^{k+1}$.

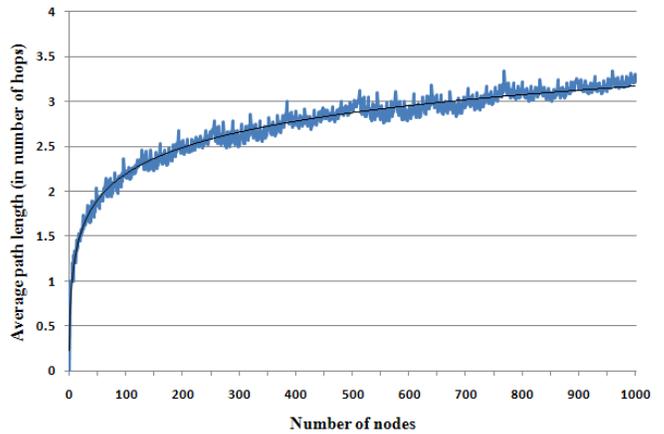

Figure 2. Average shortest path length in the proposed network (a trend line is also shown).

The second parameter we consider is the *average shortest path length*. This determines the average number of hops needed to transmit a message securely between any random pairs of nodes. Fig. 2 plots the average shortest path length for network of various sizes using the proposed model. When the network increases 10 fold from 100 nodes to 1000 nodes, the average path length increases less than 1.5 times. This shows that our model is scalable, i.e. with a very large increase in the network size the average path length only increases a small fraction.

The third parameter to consider is the *number of keys* that a node needs to store in order to ensure that the network is connected and maintain a small average path length. We know that the minimum number of connections required to ensure network connectivity is 2. That is, if a every node $i$ shares a key with node $i+1$ and $i-1$ (computed modulo $N$), then the network is connected. However, this approach results in a average shortest path length of $\frac{N}{4}$ and a diameter of $\frac{N}{2}$.

In order to reduce the path lengths, $2\log_2(N)-1$ long distance connections have been introduced, as a result, each node stores $2\log_2(N)-1$ keys, when $N$ is a power of 2. When $2^k < N < 2^{k+1}$, for some integer $k$, the number of keys vary between the number of keys for the cases $2^k$ and $2^{k+1}$.

The fourth parameter that we consider is the *clustering coefficient*. Clustering coefficient measures the local connectivity of the nodes and was introduced in [9]. It is expected, in a network, that local communication is comparatively more common than long distance communication. If nodes are deployed such that node IDs close to each other are closely located topographically, then they can communicate with each other with very few hops. Fig. 3 shows the clustering coefficient of our model when compared to a random graph.

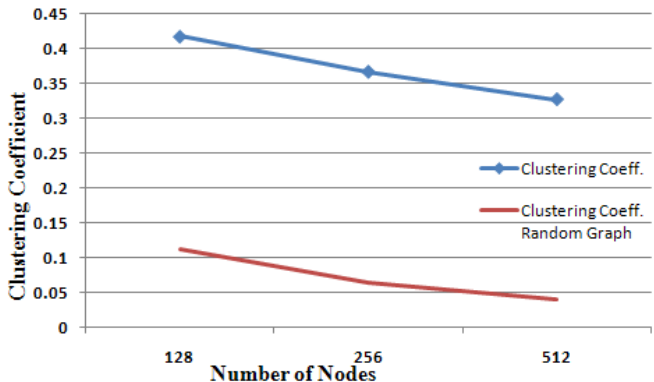

Figure 3. Clustering coefficient comparison.

**Node authentication and reuse of keys:** Since, every node connection is pre-determined and not randomly chosen, nodes know with which nodes they share keys. Similar pair-wise key distribution has been discussed in [3] and [4], however, unlike in our case, they pick the pairs randomly. Advantages of pair-wise key distribution is that we can ensure that each shared key is unique and there is no need for key discovery phase. Further, instead of randomly picking the pairs, since we have picked the pairs in a specific order, the routing procedure is predetermined.

## III. CONCLUSIONS

We have presented a scheme for key predistribution in sensor networks. The proposed scheme eliminates the key discovery and path determination overheads. We have proposed an efficient routing mechanism that guarantees $\left\lfloor \frac{\log_2 \frac{N}{8}}{2} \right\rfloor + 2$ diameter and the nodes only store $O(\log_2 N)$ keys.

Further work is needed to deal with node failures and node joins. Also, it would be of interest to determine the performance of methods based on structured graphs when only a percentage of nodes are reachable after deployment. In the case of node failures, a number of alternate routes exist since every node has $2\log_2(N)-1$ connections.

ACKNOWLEDGMENT

This research has been partly funded by the Center for Telecommunication and Network Security (CTANS), Oklahoma State University.

REFERENCES

[1] Eschenauer, L. and Gligor, V. D. 2002. A key-management scheme for distributed sensor networks. In *Proceedings of the 9th ACM Conference on Computer and Communications Security*. Pages: 18 - 22, 2002.
[2] Haowen Chan, Adrian Perrig, Dawn Song, "Random Key Predistribution Schemes for Sensor Networks," *IEEE Symposium on Security and Privacy*. Pages: 197, 2003.
[3] W. Du, J. Deng, Y. S. Han, and P. K. Varshney. A pairwise key pre-distribution scheme for wireless sensor networks. In *Proceedings of the 10th ACM Conference on Computer and Communication Security*. Pages: 42-51, 2003.
[4] D. Liu and P. Ning. Establishing pairwise keys in distributed sensor networks. In *Proceedings of the 10th ACM Conference on Computer and Communications Security*. Pages: 52-61, 2003.
[5] I. Stoica, R. Morris, D. Karger, M. F. Kaashoek and H. Balakrishnan. Chord: A scalable peer-to-peer lookup service for internet applications. In *Proceedings of the 2001 Conference on Applications, Technologies, Architectures, and Protocols For Computer Communications,* SIGCOMM '01. Pages 149-160, 2001.
[6] W. Du, J. Deng, Y.S. Han, S. Chen, and P.K. Varshney. A key management scheme for wireless sensor networks using deployment knowledge. In *Proceedings of the 23rd Annual Joint Conference of the IEEE Computer and Communications Society*, INFOCOM '04. Vol. 1, pages 586-597, 2004.
[7] D. Liu and P. Ning. Improving key predistribution with deployment knowledge in static sensor networks. *ACM Transactions on Sensor Networks*. Vol. 1. No. 2, pages 204-239, 2005.
[8] P. De, Y. Liu and S. K. Das. Deployment-aware modeling of node compromise spread in wireless sensor networks using epidemic theory. *ACM Transactions on Sensor Networks*. Volume 5, issue 3, pages 1-33, 2009.
[9] D. J. Watts and S. H. Strogatz. Collective dynamics of 'small-world' networks. Nature 393: 440-42, 1998.